\documentclass{webofc}

\usepackage[varg]{txfonts}   
\usepackage{hyperref}
\usepackage{url}
\usepackage{cleveref}
\usepackage[normalem]{ulem}
\usepackage{xcolor}
\hypersetup{colorlinks=true,citecolor=blue,urlcolor=blue,linkcolor=blue}
\newcommand\citere[1]{Ref.~\cite{#1}}
\newcommand\citeres[1]{Refs.~\cite{#1}}

\newcommand{\gev}{\;\text{GeV}\xspace}

\newcommand{\cL}{{\cal L}}
\newcommand{\cLint}{\cL_{\mathrm{int}}}

\newcommand{\vw}{\ensuremath{v_{\mathrm{w}}}}

\renewcommand{\rm}{\mathrm}
\newcommand{\nn}{\nonumber}

\newcommand{\lahhH}{\lambda_{hhH}}

\begin{document}
\begin{flushright}
\texttt{DESY-26-037}\\
\texttt{IFT–UAM/CSIC-26-026}
\end{flushright}

\title{Investigating a strong first-order electroweak phase transition in the RxSM at future linear $e^+e^-$ colliders and LISA}

\author{\firstname{Johannes} \lastname{Braathen}\inst{1}\fnsep\thanks{\email{johannes.braathen@desy.de}, \textit{Speaker}} \and
        \firstname{Sven} \lastname{Heinemeyer}\inst{2}\fnsep\thanks{\email{sven.heinemeyer@cern.ch}} \and
        \firstname{Carlos} \lastname{Pulido Boatella}\inst{2}\fnsep\thanks{\email{carlos.pulido@estudiante.uam.es}} \and
        \firstname{Alain} \lastname{Verduras Schaeidt}\inst{1}\fnsep\thanks{\email{alain.verduras@desy.de}}
}

\institute{Deutsches Elektronen-Synchrotron DESY, Notkestr.~85,  22607  Hamburg,  Germany
\and
Instituto de F\'isica Te\'orica (UAM/CSIC), 
Universidad Aut\'onoma de Madrid, Cantoblanco, 28049, Madrid, Spain
          }

\abstract{
The general real singlet extension of the Standard Model (SM), the RxSM, is one of the simplest theories Beyond-the-Standard Model (BSM) that can accommodate a strong first-order electroweak phase transition (SFOEWPT). We investigate the possible thermal histories of the scalar potential in the RxSM, and the regions of the model parameter space in which SFOEWPT can be realised. We then explore complementary avenues to probe such scenarios experimentally: either using searches for a stochastic background of gravitational waves (GWs), or using searches for di-Higgs production processes at future collider experiments, focusing on the case of a high-energy $e^+e^-$ collider.
An important aspect of our work is that one-loop corrections to all relevant trilinear scalar couplings are consistently included both in the calculation of dynamics of the electroweak phase transition (EWPT) and in collider processes. 
We find entirely different phenomenological signatures for different parts of the RxSM parameter space giving rise to SFOEWPTs. 
On the one hand, if the SFOEWPT is driven by the singlet field, the 125 GeV Higgs boson is very SM-like and signs of BSM physics would be difficult to find at colliders, but strong GW signals could be produced. On the other hand, in scenarios where a SFOEWPT is driven by the doublet field, BSM deviations in properties of the detected Higgs boson, particularly in its trilinear self-coupling, typically lead to observable signals at colliders, while detectable GW signals are much more challenging to achieve.   
This work highlights the complementarity of collider experiments and cosmological observations to determine the dynamics of the EWPT and reconstruct the shape of the Higgs potential realised in Nature. 
}
\setlength{\footskip}{20pt}
\pagestyle{plain}
\maketitle
\vfill
\noindent\textit{Talk presented at the International Workshop on Future Linear Colliders 2025 (LCWS2025), Valencia, Spain}
\newpage
\section{Introduction}
\label{intro}
The origin of the observed baryon asymmetry of the Universe (BAU)~\cite{Planck:2018vyg,Fields:2019pfx} is one of the most pressing questions left unanswered by our current understanding of High-Energy Physics. Among the potential  scenarios to explain the BAU, electroweak baryogenesis (EWBG)~\cite{Kuzmin:1985mm,Cohen:1993nk} is especially attractive as it only involves electroweak scale phenomena and is therefore testable in the foreseeable future. 
In this context, the possibility of the electroweak phase transition (EWPT) occuring as a strong first-order electroweak phase transition (SFOEWPT) is of particular relevance: a SFOEWPT would lead to departure from thermal equilibrium, one of the Sakharov conditions~\cite{Sakharov:1967dj} that need to be fulfilled for a dynamical generation of the BAU via EWBG.

An important task for particle physics is thus to investigate possible experimental probes of SFOEWPTs. A first option is to reconstruct the shape of the Higgs potential that is realised in Nature. This can be done by accessing Higgs pair production processes, among which $gg\to hh$ at the LHC and its high-luminosity upgrade (HL-LHC), or $e^+e^-\to Zhh$ and $e^+e^-\to \nu\bar{\nu}hh$ at high-energy $e^+e^-$ colliders~\cite{Barklow:2017awn,LinearColliderVision:2025hlt,Altmann:2025feg}. These provide in particular direct access to the trilinear Higgs self-coupling, $\lambda_{hhh}$, that controls the shape of the Higgs potential away from the EW minimum, along the Higgs direction in field space. In particular, a SFOEWPT occurring along the Higgs field direction would be correlated with a significant deviation in $\lambda_{hhh}$ from its SM prediction --- see e.g.\ Refs.~\cite{Grojean:2004xa,Kanemura:2004ch,Kakizaki:2015wua,Hashino:2016rvx,Hashino:2016xoj,Basler:2017uxn,Biekotter:2022kgf,Bittar:2025lcr,Braathen:2025svl}. In models with extended scalar sectors, large BSM deviations in $\lambda_{hhh}$ can arise due to radiative corrections from the BSM scalars~\cite{Kanemura:2002vm,Kanemura:2004mg,Aoki:2012jj,Kanemura:2015fra,Kanemura:2015mxa,Arhrib:2015hoa,Kanemura:2016sos,Kanemura:2016lkz,He:2016sqr,Kanemura:2017wtm,Kanemura:2017gbi,Chiang:2018xpl,Basler:2018cwe,Senaha:2018xek,Braathen:2019pxr,Kanemura:2019slf,Braathen:2019zoh,Braathen:2020vwo,Basler:2020nrq,Bahl:2022jnx,Bahl:2022gqg,Falaki:2023tyd,Bahl:2023eau,Aiko:2023nqj,Cherchiglia:2024abx,Basler:2024aaf,Bahl:2025wzj,Arco:2025pgx,Braathen:2025qxf}, even in scenarios where other properties of the 125 GeV Higgs boson are SM-like, e.g.\ in the so-called alignment limit~\cite{Gunion:2002zf} of BSM models. 

An alternative approach is to search for cosmological relics of a SFOEWPT in the early Universe; see e.g.\ \citere{Athron:2023xlk} for a review. These could include a stochastic background of gravitational waves (GWs)~\cite{Grojean:2006bp,Ashoorioon:2009nf,No:2011fi,Huber:2015znp,Kakizaki:2015wua,Hashino:2016rvx,Dorsch:2016nrg,Kang:2017mkl,Bruggisser:2018mrt,Chala:2018ari,Morais:2018uou,Hashino:2018zsi,Hashino:2018wee,Goncalves:2021egx,Biekotter:2022kgf,Lewicki:2024ghw,Biekotter:2025npc}, primordial black holes~\cite{Kodama:1982sf,Liu:2021svg,Hashino:2021qoq,Jung:2021mku,Kawana:2022olo,Lewicki:2023ioy,Gouttenoire:2023naa,Baldes:2023rqv,Flores:2024lng,Lewicki:2024ghw,Kanemura:2024pae,Hashino:2025fse,Franciolini:2025ztf,Kierkla:2025vwp} or primordial magnetic fields~\cite{Vachaspati:1991nm,Ellis:2019tjf,Olea-Romacho:2023rhh}. In particular, GWs produced during a SFOEWPT by collisions of the nucleating bubbles of the true vacuum or by sound waves and later turbulence in the plasma around these bubbles would have a frequency range peaking within the range of sensitivity of space-based GW interferometers like LISA~\cite{Caprini:2019egz,LISACosmologyWorkingGroup:2022jok}.  

While the Standard Model (SM) cannot accommodate a SFOEWPT, simple extensions of the SM Higgs sector make this possible. 
The general real singlet extension of the SM (RxSM) is one of the simplest models such models, yet it still offers a rich phenomenology. 
Numerous papers have investigated the dynamics of the EWPT in the RxSM (see e.g.\ Refs.~\cite{Espinosa:2007qk,Profumo:2007wc,Huang:2016cjm,Alves:2018jsw,Liu:2021jyc,Ellis:2022lft,Blasi:2023rqi,Ramsey-Musolf:2024ykk,Niemi:2024vzw,Gould:2024jjt,Niemi:2024axp,Ghosh:2022fzp,Roy:2022gop,Goncalves:2024vkj}), often in relation with collider signatures of the model~\cite{Li:2019tfd,Liu:2021jyc,Zhang:2023jvh,Palit:2023dvs,Ramsey-Musolf:2024ykk} (see also Refs.~\cite{Feuerstake:2024uxs,Lewis:2024yvj,Aboudonia:2024frg}). In Ref.~\cite{Braathen:2025svl}, we have explored the interplay between Higgs pair production processes at the HL-LHC and at $e^+e^-$ colliders with searches for gravitational waves at LISA to probe the thermal history of the RxSM scalar potential. The investigation of the early-Universe evolution of the scalar potential is performed at the one-loop level, and in order to ensure consistency with the collider analysis, we also incorporate full one-loop corrections to the trilinear scalar couplings relevant for di-Higgs production processes, using our results and renormalisation scheme from Ref.~\cite{Braathen:2025qxf}.

In these proceedings, we summarise our findings in Ref.~\cite{Braathen:2025svl}, with a particular focus on the potential of high-energy $e^+e^-$ colliders to probe scenarios with a SFOEWPT. 


\section{Model and analysis setup}

We consider in our study the general singlet extension of the SM, or RxSM. We refer the reader to Refs.~\cite{Braathen:2025qxf,Braathen:2025svl} for details on our notations and conventions, and we only recall here the tree-level scalar potential of the model
\begin{align}
    V^{(0)}({\Phi},{ S})=\mu^2|\Phi|^2+\frac{\lambda}{2}|\Phi|^4+\kappa_{SH}|\Phi|^2 S+\frac{\lambda_{SH}}{2}|\Phi|^2 S^2+\frac{M_S^{2}}{2}{S}^2+\frac{\kappa_S}{3}{ S}^3+\frac{\lambda_S}{2}S^4\,, \label{eq:potential}
\end{align}
where $\Phi$ is the Higgs doublet, $S$ is the additional real singlet, and all parameters are real.  
After mixing, the scalar sector of the RxSM consists of two states, $h$ and $H$, of which we consider that the lighter one, $h$, corresponds to the detected Higgs boson. The scalar sector is parametrised in terms of seven parameters: the masses of the two CP-even states $m_h$ and $m_H$, their mixing angle $\alpha$, the vacuum expectation values (VEVs) of the doublet, $v$, and of the singlet, $v_S$, as well as two Lagrangian trilinear scalar couplings $\kappa_S$ and $\kappa_{SH}$ (which correspond to terms $\mathcal{L}\supset -\kappa_S S^3/3-\kappa_{SH}|\Phi|^2S$ in the Lagrangian, see e.g.\ Eq.~(2) of Ref.~\cite{Braathen:2025svl}). 

We include in our analysis state-of-the-art theory constraints on the RxSM parameter space, in particular  boundedness-from-below of the potential, perturbativity~\cite{Li:2019tfd}, and perturbative unitarity~\cite{Braathen:2017jvs}. On the experimental side, we also verify using \texttt{HiggsTools}~\cite{Bahl:2022igd} (see \citeres{Bechtle:2008jh,Bechtle:2011sb,Bechtle:2013wla,Bechtle:2015pma,Bechtle:2020pkv,Bechtle:2013xfa,Bechtle:2014ewa,Bechtle:2020uwn} for details)
that the properties of $h$ are compatible with that of the detected Higgs boson, and that direct searches for BSM scalars do not exclude the points we consider. 

To analyse the dynamics of the EWPT, we employ the public tool \texttt{BSMPT}~\cite{Basler:2018cwe,Basler:2020nrq,Basler:2024aaf}, version \texttt{3.1.1}, in which we have implemented the RxSM~\cite{Braathen:2025svl}. \texttt{BSMPT} allows to trace the evolution of the vacuum phases of the theory with temperature, and to determine thermodynamical quantities that describe the dynamics of the EWPT. Specifically, we consider 
\begin{itemize}
    \item the nucleation temperature $T_n$ and the percolation temperature $T_p$;
    \item $\xi_n\equiv v(T_n)/T_n$ , which quantifies the strength of the phase transition;
    \item $\alpha$, which is the latent heat released during the transition; and
    \item $\beta/H_*$, which is the inverse duration of the transition (in units of the Hubble rate at the time of the EWPT).
\end{itemize}
We also use \texttt{BSMPT} --- which itself makes use of the results of Ref.~\cite{Caprini:2019egz} --- in order to compute the spectrum of gravitational waves produced during the SFOEWPT and the signal-to-noise ratio (SNR) that would be obtained at the LISA space-based interferometer. 

This RxSM features two trilinear scalar couplings relevant for pair production processes of the detected Higgs boson, namely $\lambda_{hhh}$ and $\lambda_{hhH}$. These are computed at the complete one-loop level using the public tool \texttt{anyH3}~\cite{Bahl:2023eau,Bahl:2026anyhh,Bahl:2026anyhhproc}, using the fully on-shell renormalisation scheme devised in Ref.~\cite{Braathen:2025qxf}. 
In turn, for the di-Higgs production processes $e^+e^-\to Zhh$ and $e^+e^-\to \nu\nu hh$, we obtain predictions for total cross-sections and differential $m_{hh}$ distributions employing \texttt{Madgraph5\_aMC  v3.5.9}~\cite{Alwall:2014hca}, taking as considered collider setup the ILC1000~\cite{ILC:2013jhg,Moortgat-Pick:2015lbx,Bambade:2019fyw} (i.e. $\sqrt{s}=1\text{ TeV}$). The \texttt{UFO} model files for the RxSM were created using \texttt{SARAH v4.15}~\cite{Staub:2008uz,Staub:2009bi,Staub:2010jh,Staub:2012pb,Staub:2013tta,Staub:2015kfa}. 


\section{Dynamics of the EWPT in the RxSM}

\begin{figure}[ht]
    \centering
    \includegraphics[width=0.32\linewidth]{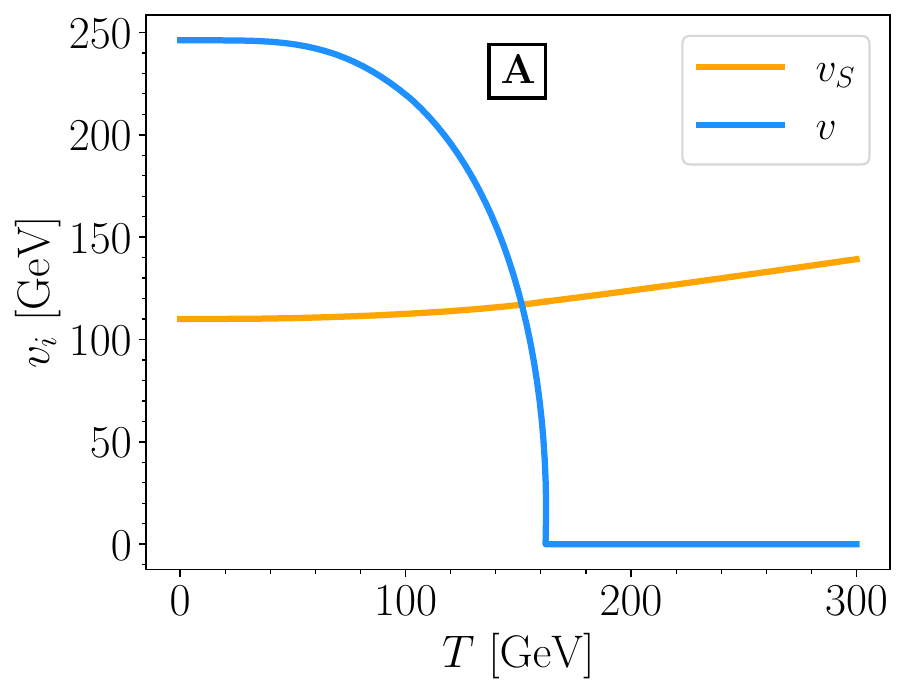}
    \includegraphics[width=0.32\linewidth]{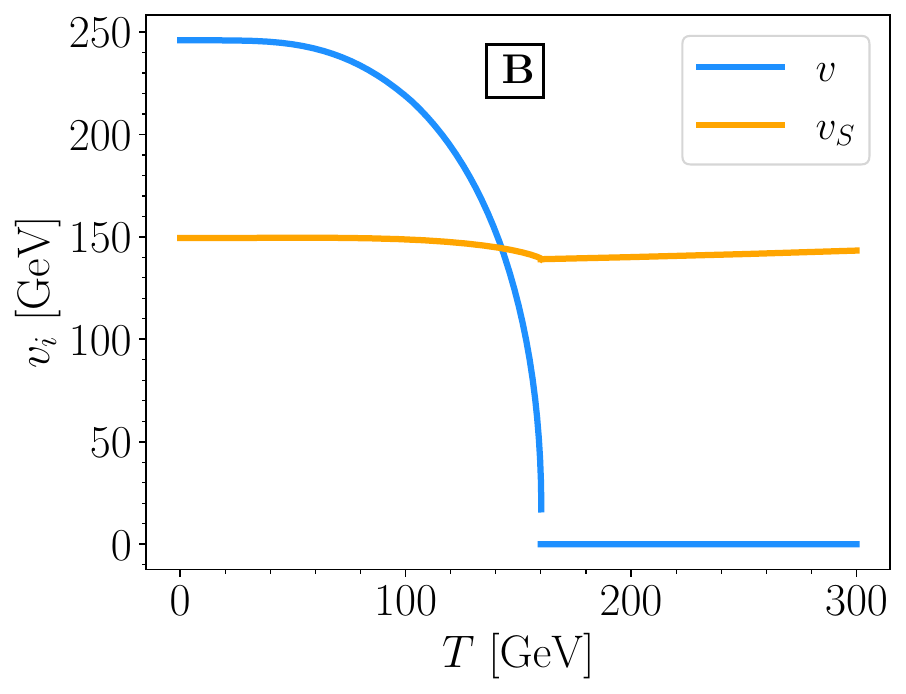}
    \includegraphics[width=0.32\linewidth]{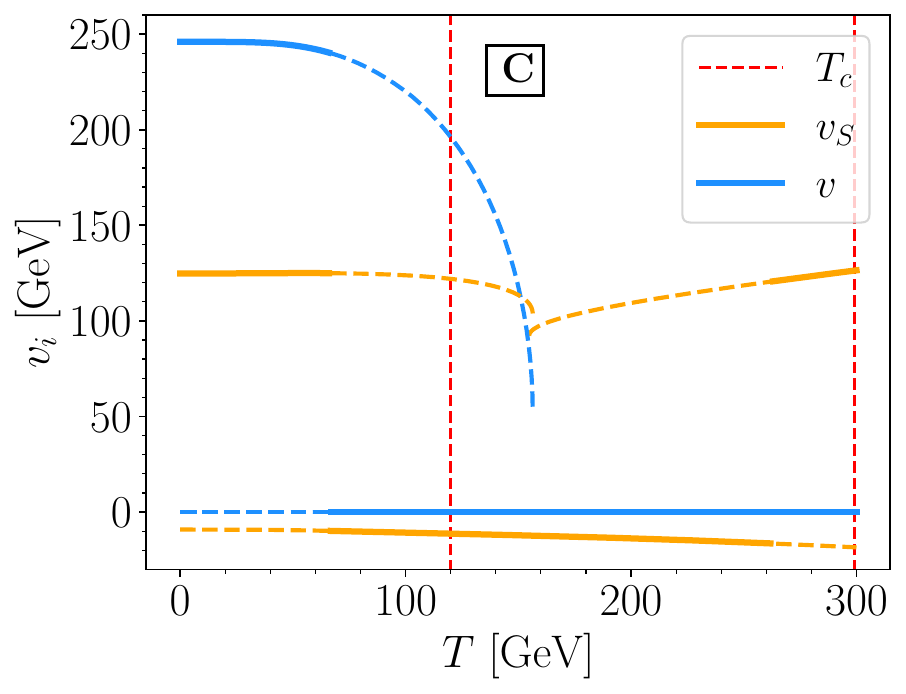}
    \includegraphics[width=0.32\linewidth]{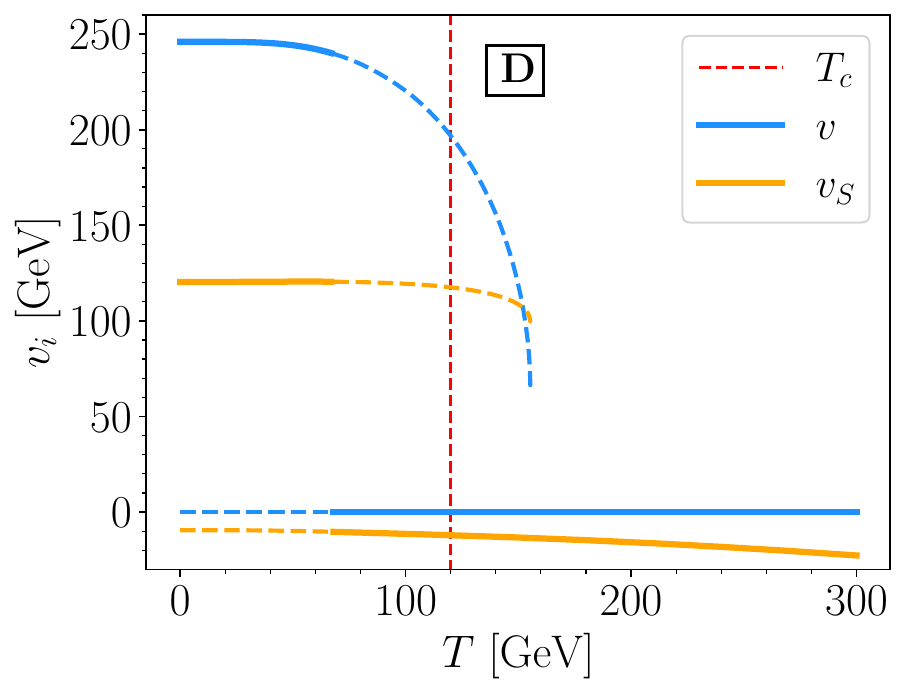}
    \includegraphics[width=0.32\linewidth]{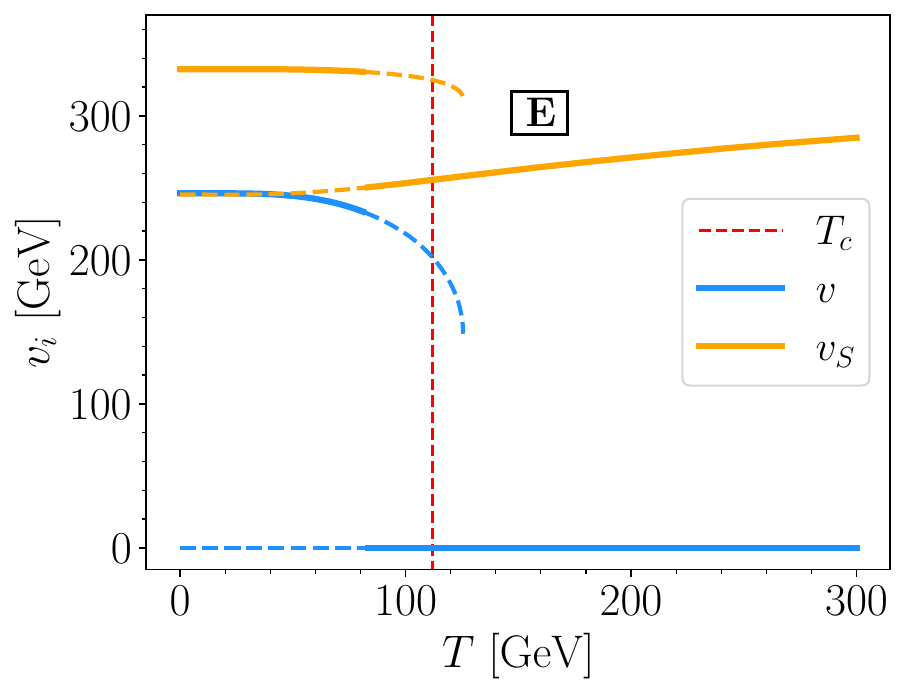}
    \includegraphics[width=0.32\linewidth]{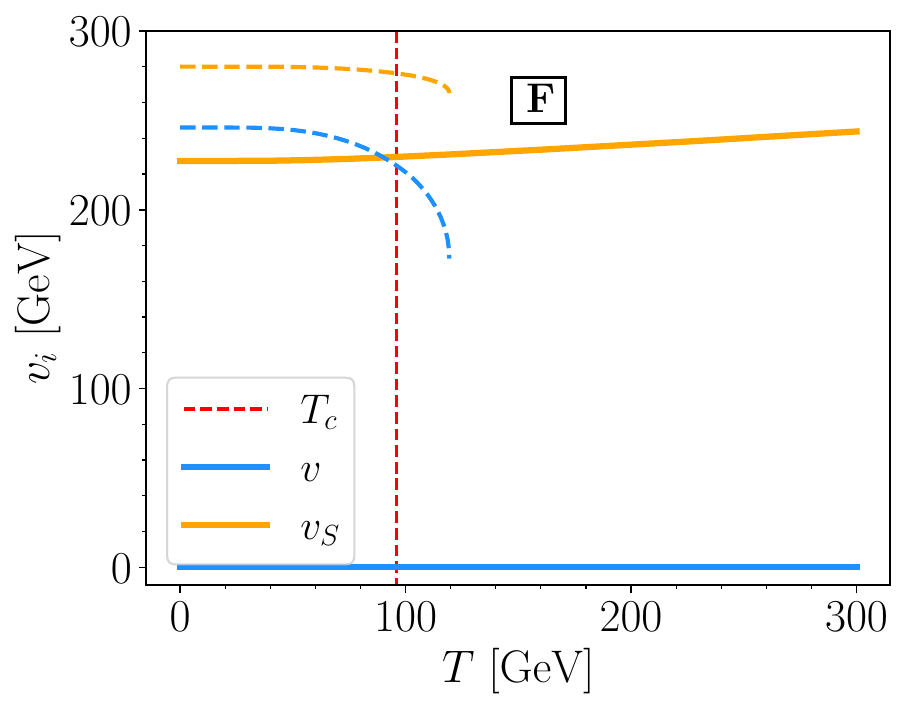}
    \caption{Tracing of the different minima of the potential as a function of the temperature for the six thermal histories possible in the RxSM. Blue lines represent the minima of the EW doublet and orange lines those of the singlet field, while red vertical lines indicate the critical temperature $T_c$. The solid lines represent the path followed by the Universe.}
    \label{fig:thermalhistories}
\end{figure}

We begin by exploring the possible thermal histories of the early Universe in the RxSM. We have performed an extensive scan of the model parameter space, and determined the thermal evolutions of the minima of the potential for points allowed by all current experimental and theoretical constraints, but {\em not} requiring a SFOEWPT. Among our scan points we have found six possible types of behaviours (``thermal histories''), illustrated in \cref{fig:thermalhistories}. Specially, we have   
traced for each case the evolution of the minima of the potential as a function of the temperature. The trajectories of the doublet and singlet VEVs are shown in blue and orange, respectively, and the solid lines indicate the vacuum phases of the doublet and singlet in which the Universe actually is during the temperature evolution. 

Among the six cases, three are of particular interest as they correspond to SFOEWPTs: case \textbf{C} is a two-step SFOEWPT, in which there is a first (first-order) transition only along the singlet field direction, followed by a second (also first-order) transition along the doublet and singlet directions simultaneously. Meanwhile cases \textbf{D} and \textbf{E} are two types of single-step SFOEWPT, differing by whether the initial (high-temperature) value of $v_S$ is negative (case \textbf{D}) or positive (case \textbf{E}). The difference in the high-temperature value of $v_S$ has a strong impact on which field dominates  dynamics of the EWPT and in turn on the associated phenomenology, as we will see below. 
Concerning the other three cases: cases \textbf{A} and \textbf{B} are respectively a second-order EWPT and a first-order EWPT so weak that it is phenomenologically not distinguishable from \textbf{A}. Case \textbf{F} corresponds to vacuum trapping~\cite{Cline:1999wi,Baum:2020vfl,Biekotter:2021ysx,Biekotter:2022kgf}, i.e.\ the unphysical case where the vacuum remains in the unbroken phase of EW symmetry and the tunnelling rate to the EW vacuum is longer than the age of the Universe. 

\begin{figure}
    \centering
    \includegraphics[width=0.49\textwidth]{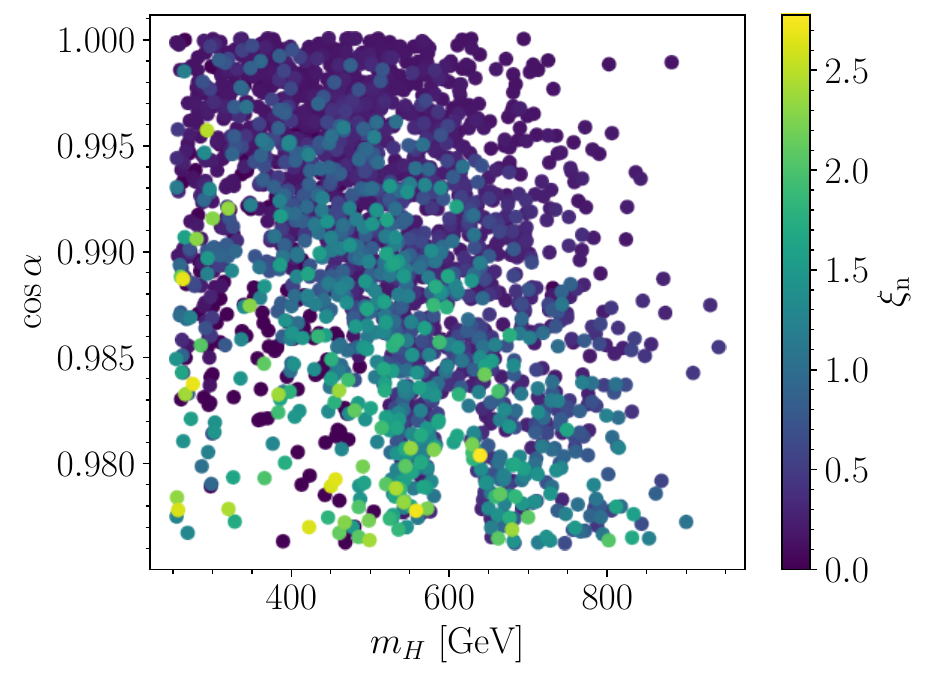}
    \includegraphics[width=0.48\textwidth]{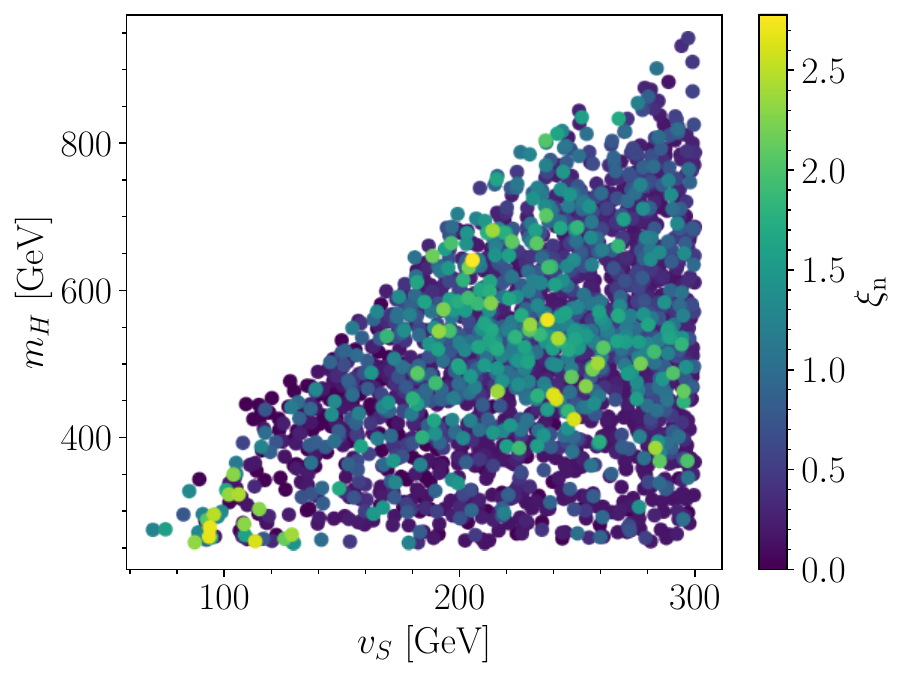}
    \caption{RxSM parameter scan results for $\xi_n\equiv v_n/T_n$. 
    \textit{Left:} $\{m_H,\cos \alpha\}$ plane; \textit{right}: 
    $\{v_S,m_H\}$ plane.}
    \label{fig:scatter}
\end{figure}

As a next step, we have computed for the scan points the strength of the EWPT, $\xi_n$, in order to identify the regions of the RxSM parameter space in which SFOEWPTs are possible. The corresponding results are shown in \cref{fig:scatter} in the two phenomenologically most relevant  planes $\{m_H,\cos\alpha\}$ (left) and $\{v_S,m_H\}$ (right). Points with $\xi_n>1$ (and with values reaching 2.5 and more) are found in two regions, namely:
\begin{itemize}
    \item a first region with $v_S\lesssim 150\text{ GeV}$ and $m_H\lesssim 400 \text{ GeV}$ (as well as $\kappa_{SH}> -600\text{ GeV}$ and $\kappa_S>-300 \text{ GeV}$, c.f.\ Ref.~\cite{Braathen:2025svl}); and
    \item a second region characterised by $v_S\gtrsim 200\text{ GeV}$ and $m_H\gtrsim 400\text{ GeV}$ (as well as $\kappa_{SH}<-500\text{ GeV}$~\cite{Braathen:2025svl}).
\end{itemize}
In the following we define for each of these two regions a benchmark plane that will serve to investigate further the dynamics of the EWPT. 


\section{Benchmark plane 1: a singlet-driven EWPT}

The first benchmark plane (BP1), related to the first region, is defined by the following ranges and relation between parameters
\begin{align}
    m_H&\in[260,1000]~\mathrm{GeV}\,,\ 
    \cos\alpha\in[0.975,1]\,,\ \kappa_S=-900~\mathrm{GeV}\,,\nn\\
    \kappa_{SH}&=(5662.9\cos\alpha-5688.4)~\mathrm{GeV}\,,\ 
    v_S=(4239.5\cos\alpha-4067.6)~\mathrm{GeV}\,.
    \label{bp1}
\end{align}

\begin{figure}
    \centering
    \includegraphics[width=0.65\textwidth]{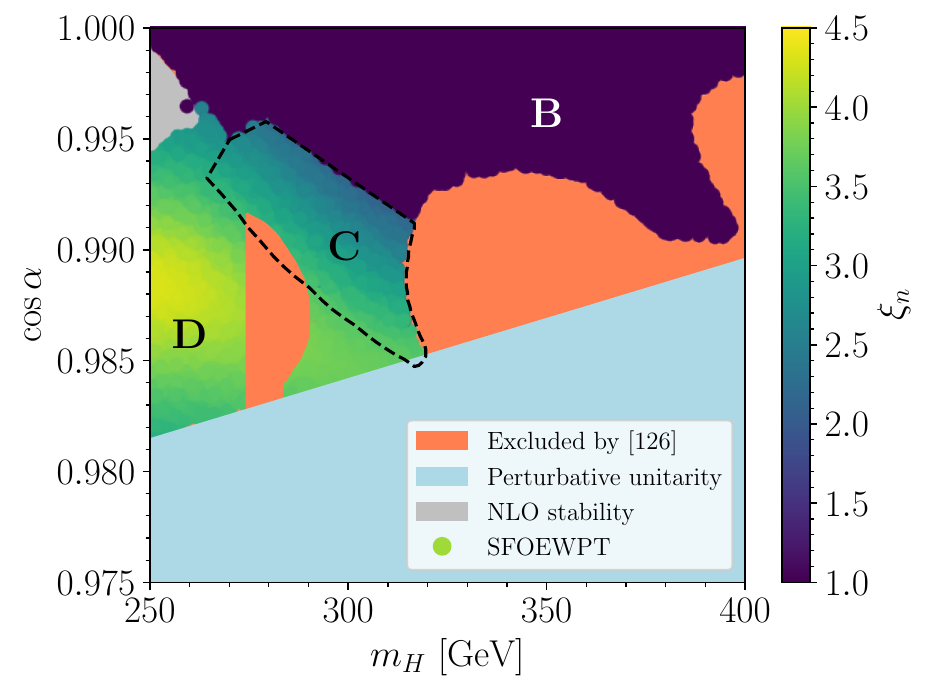}
    \caption{Parameter scan results for BP1, with $\xi_n$ indicated by the colour of the scatter points. 
    The different thermal histories are labelled following \cref{fig:thermalhistories}. Coloured regions are excluded perturbative unitarity (light blue), NLO stability of the EW vacuum (grey), and direct searches for heavy Higgs bosons~\cite{ATLAS:2020tlo} (light red). In region \textbf{B}, $\xi_n\in [0,1]$.
    }
    \label{fig:BP1diags}
\end{figure}

In \cref{fig:BP1diags} we present the dynamics of the EWPT and the value of $\xi_n$ (indicated by the colour coding) in the parts of BP1 allowed by theoretical and experimental constraints. The light blue and grey regions are excluded by perturbative unitarity and stability of the scalar potential at one loop, respectively, while the orange area is excluded by direct searches for a BSM Higgs boson decaying into a pair of $Z$ bosons~\cite{ATLAS:2020tlo}.  Close to the alignment limit ($\cos\alpha\to 1$), a significant part of the plane (shown in dark blue) features a first-order EWPT, that is however not strong enough ($\xi_n<1$, region \textbf{B}) to be of phenomenological relevance in connection to the explanation of the BAU. On the other hand, for $\cos\alpha\lesssim 0.996$ and lower values of $m_H$ we find regions with SFOEWPTs proceeding either in two steps (case \textbf{C}, within the black dashed line) or in a single step (case \textbf{D}) --- for convenience we name the regions with the same letter as the corresponding case from \cref{fig:thermalhistories}. The smooth transition from region \textbf{C} to \textbf{D} can be understood from the continuous deformation of the second minimum of the singlet field at high temperatures. 

\begin{figure}
    \centering
    \includegraphics[width=0.48\textwidth]{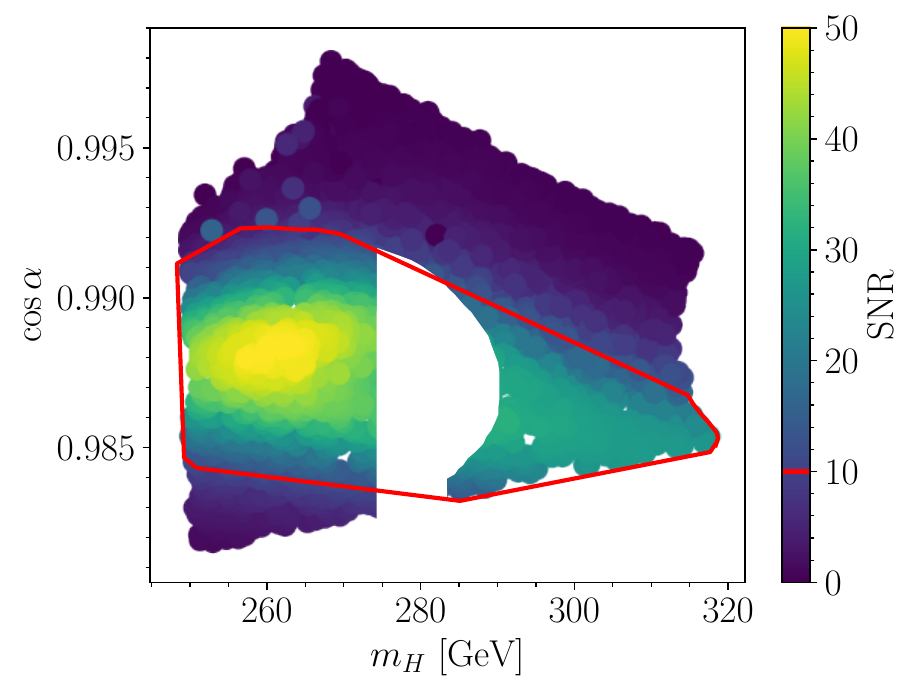}
    \includegraphics[width=0.5\textwidth]{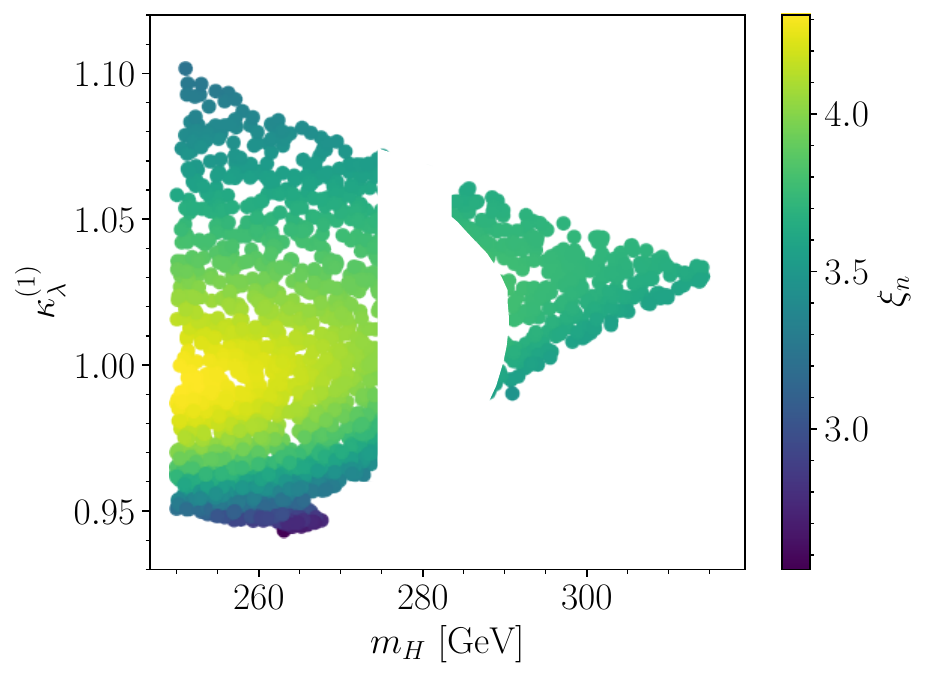}
    \caption{Parameter scan results for points in benchmark plane 1 with a SFOEWPT (regions \textbf{C} and \textbf{D} of \cref{fig:BP1diags}). \textit{Left:} SNR at LISA (assuming $\vw=0.95$ and three years of observation time) in the plane $\{m_H,\cos\alpha\}$. The red line indicates the region with $\mathrm{SNR} > 10$. \textit{Right:} $\xi_n$ in the plane $\{m_H,\kappa_\lambda^{(1)}\}$ for points with $\mathrm{SNR} > 10$.}
    \label{fig:BP1results}
\end{figure}

Next, in \cref{fig:BP1results}, we display the scan points in the most interesting regions \textbf{C} and \textbf{D}, indicating  phenomenologically relevant quantities describing potential probes of the scenarios in BP1 via gravitational waves signals (left) or at colliders (right). Specifically, the left panel of \cref{fig:BP1results} shows the SNR at LISA in the plane $\{m_H,\cos\alpha\}$, taking the pessimistic assumption\footnote{In various studies of theories with extended Higgs sectors~\cite{Ai:2023see,Ekstedt:2024fyq,Ellis:2022lft}, the bubble wall velocity was found to be in the range $\vw\in[0.2,1]$, and in Ref.~\cite{Braathen:2025svl} we showed that in the RxSM, taking $\vw=0.95$ leads to conservative estimates of the SNR (see also Refs.~\cite{Branchina:2024rva, DeCurtis:2024hvh, DeCurtis:2023hil, DeCurtis:2022hlx,Branchina:2025adj} for more detailed studies of the bubble wall dynamics).} of $\vw=0.95$ for the velocity of bubble walls during the SFOEWPT and considering three years of observation time. We note that in case \textbf{C} GWs can be sourced by both parts of the two-step phase transition. However, the GW signal produced by the first transition, occurring exclusively along the singlet-field direction, was found to be much smaller and negligible compared to the signal arising from the second transition that also involves the doublet-field direction. Therefore, the SNR values shown here correspond to the second transition only (while the contribution of the first, singlet-like, transition is subleading).  
The red line indicates the region with an $\mathrm{SNR} > 10$.  
Here it should be kept in mind that the choice of $\kappa_{SH}$ and $v_S$ that defines BP1, see \cref{bp1} was made to ensure that large parts of this parameter plane exhibit high values of the SNR (the maximal values of the SNR that are reached are however not particularly dependent on this choice). The locations for which the largest SNR values are obtained can be related to the thermodynamical quantities describing the EWPT, in particular the released latent heat $\alpha$. For BP1, the parameters computed by \texttt{BSMPTv3} are in the ranges $\alpha\in[0,0.4]$, $\beta/H_\star\in[0,1050]$, and $T_\star\in[49,100]$ GeV; and the largest SNR are associated with the highest values of $\alpha$, as well as low values of $\beta/H$ and $T_\star$. 
On the other hand, the right panel of \cref{fig:BP1results}~displays the one-loop result for the coupling modifier of the trilinear Higgs self-coupling, $\kappa_\lambda\equiv\lambda_{hhh}/\lambda_{hhh}^{\text{SM},\ (0)}$ (where $\lambda_{hhh}^{\text{SM},\ (0)}$ is the tree-level SM prediction), as a function of $m_H$ (and the colour coding additionally indicates $\xi_n$). Here we focus on the (significant) part of BP1  with $\mathrm{SNR} > 10$, i.e.\ giving rise to a stochastic background  of gravitational waves detectable at LISA (corresponding to the red contour line in the left plot). We find that no large deviations occur in the trilinear Higgs self-coupling (nor in other Higgs couplings). In turn, this implies that collider probes of scenarios like BP1 appear challenging.


\section{Benchmark plane 2: a doublet-driven EWPT}

We turn now to the second plane (BP2), related to the second region found in \cref{fig:scatter}, which is defined by 
\begin{align}
    \cos \alpha &= 0.98\, ,\ \kappa_S = -300\text{ GeV}\, ,\ v_S = 280\text{ GeV}\,,\ m_H\in [260,1000]\text{ GeV}\,, \nn\\
    \kappa_{SH} &\in [-2, -1] \text{ TeV}\, .
\label{eq:defBP2}
\end{align}

\begin{figure}[ht!]
    \centering
    \includegraphics[width=0.5\textwidth]{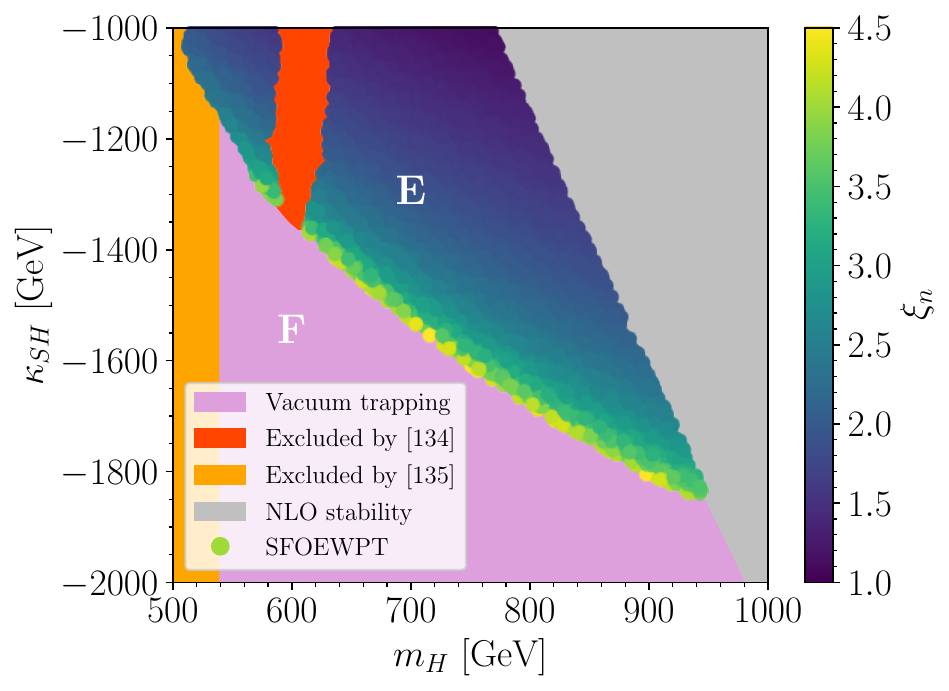}
    \includegraphics[width=0.47\textwidth]{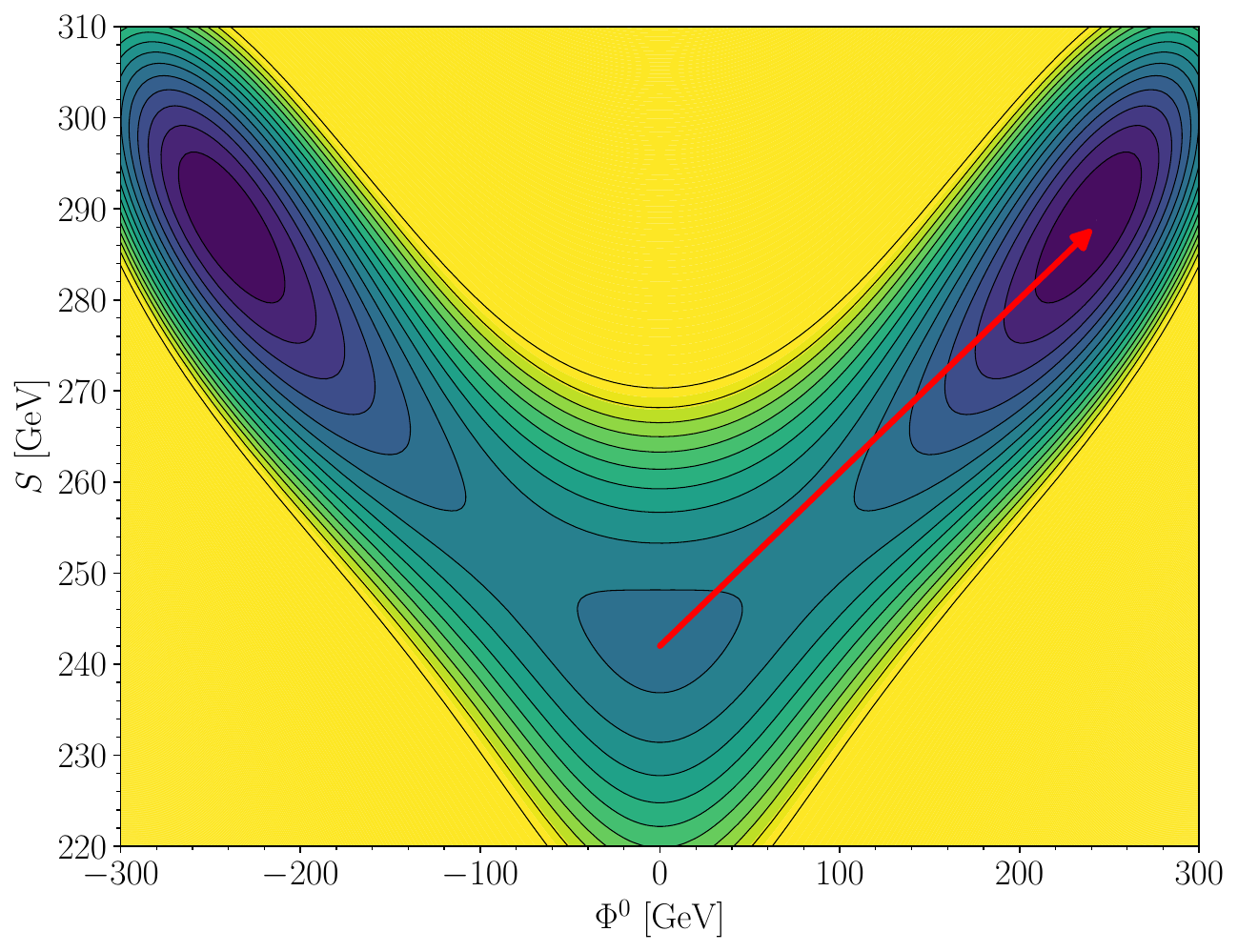}
    \caption{\textit{Left:} parameter scan results for benchmark plane 2, with $\xi_n$ shown by the colour coding of the scatter points. The labelling of different thermal histories follows that of \protect\cref{fig:thermalhistories}.  Coloured regions are excluded by vacuum trapping (pink), NLO stability of the potential (grey), or experimental searches (red by Ref.~\cite{ATLAS:2018sbw}, and orange by Ref.~\cite{CMS:2021yci}).  \textit{Right:} Temperature-dependent one-loop effective potential at the nucleation temperature, $V_\text{eff}(T_n)$, for a benchmark point with a thermal history of type \textbf{E}. This point is defined by $m_H=639 \gev$, $c_{\alpha}=0.9777$, $v_S=289 \gev$, $\kappa_S=-205 \gev$, $\kappa_{SH}=-1403 \gev$, where we find $T_n=64\gev$ and a $\mathrm{SNR} = 4.5$. Here $S$ and $\Phi^0$ denote the singlet field and the CP-even neutral component of the doublet, respectively.}
    \label{fig:BP2diags}
\end{figure}

The dynamics of the EWPT in the allowed parts of this plane are shown in the left plot of \cref{fig:BP2diags}. The grey shaded area is excluded by the theoretical constraint of stability of the potential (at one loop), while the red and orange regions are excluded by the direct searches for BSM scalars in \citere{ATLAS:2018sbw} (red) and in \citere{CMS:2021yci} (orange). The remainder of the plane is in principle allowed by theoretical constraints and experimental searches at colliders, however, the pink coloured area is excluded as these points feature vacuum trapping, case~\textbf{F} in \cref{fig:thermalhistories} and are thus not phenomenologically viable.

The right plot of \cref{fig:BP2diags} provides an illustration of the full one-loop effective potential computed at the nucleation temperature $V_\text{eff}(T_n)$, for a benchmark point featuring a SFOEWPT with SNR\;$=4.5$ and $T_n=64\gev$ (with the same type of dynamics as points from BP2, and the parameters are given in the legend). 
In this scenario, the singlet field direction does not play a major role in the EWPT, and the difference between the initial and final values of $v_S$ is small. On the other hand, the EW doublet direction is the most important one as the tunnelling distance is much larger. We can therefore distinguish the situation in BP2, where the EW doublet direction plays the most important role at the time of the transition, with that of BP1, where it is instead the singlet field direction. This finding also helps understand the behaviour observed for the transition between SFOEWPT and vacuum trapping in \cref{fig:BP2diags}. This is similar to the case of the 2HDM~\cite{Biekotter:2022kgf}, where the EWPT is driven by the EW VEV. On the contrary, in benchmark plane 1 the singlet phase plays a more important role and dynamics of the EWPT significantly differ from those in the 2HDM, as discussed above.

\begin{figure}
    \centering
    \includegraphics[width=0.5\textwidth]{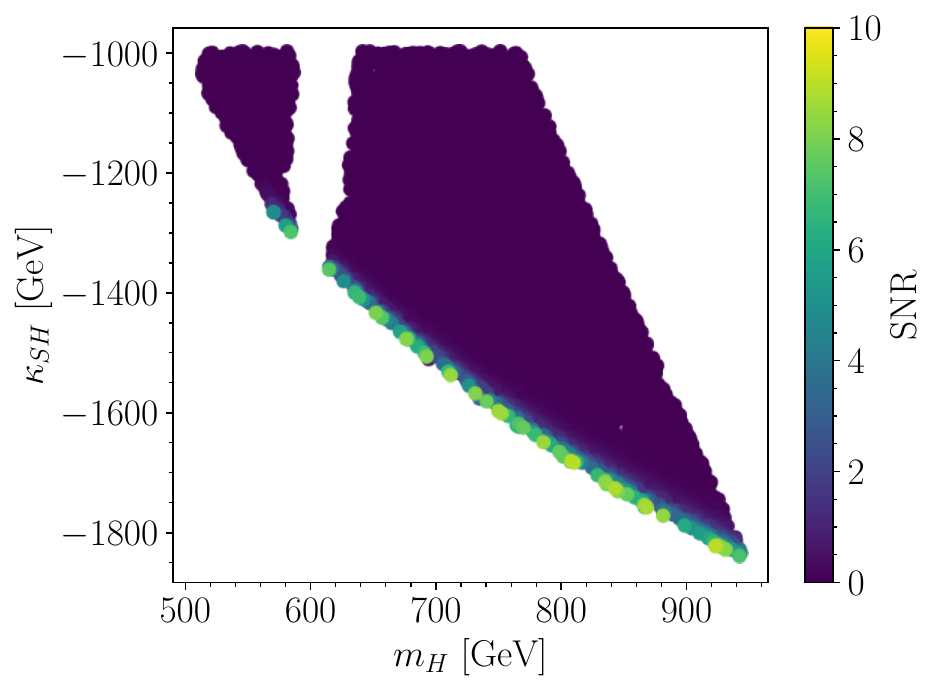}
    \includegraphics[width=0.48\textwidth]{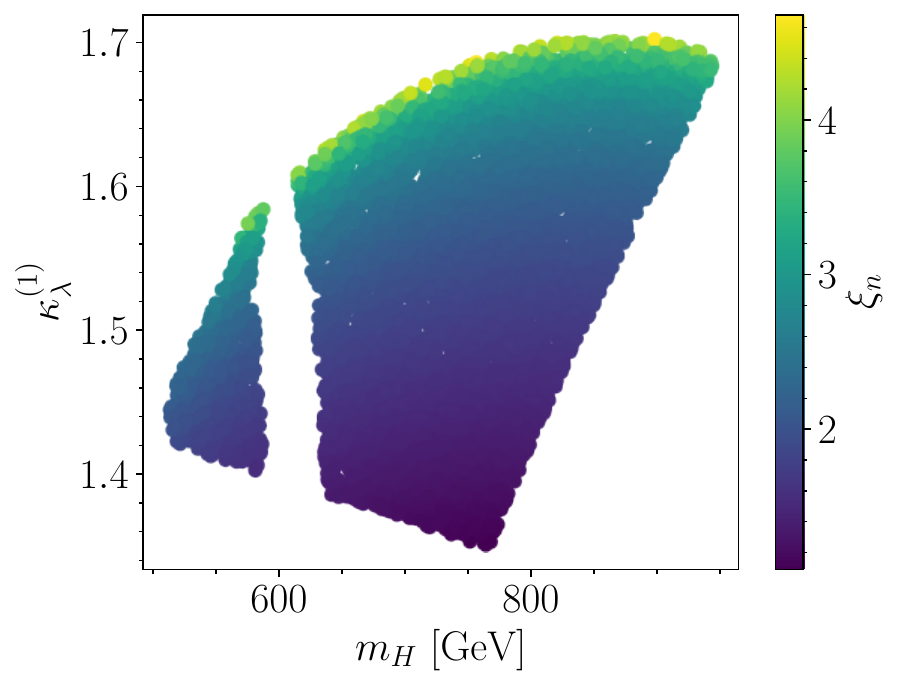}
    \caption{Parameter scan results for points in benchmark plane 2 with a SFOEWPT (region \textbf{E} of \cref{fig:BP2diags}). \textit{Left:} SNR at LISA (assuming $\vw=0.95$ and three years of observation time) in the plane $\{m_H,\kappa_{SH}\}$; \textit{right:} $\xi_n$ in the plane $\{m_H,\kappa_\lambda^{(1)}\}$.}
    \label{fig:BP2results}
\end{figure}

The left plot of \cref{fig:BP2results} displays for the region~\textbf{E} of BP2 the SNR of a potential gravitational wave signal at LISA (assuming a pessimistic value of $\vw=0.95$ for the wall velocity and three years of data taking), projected in the plane $\{m_H,\kappa_{SH}\}$. As can be seen, no point in BP2 exceeds an SNR of 10, and only a narrow band of points, near the region of vacuum trapping reaches $\text{SNR}\gtrsim 8$. The actual observability of these points will eventually depend on the true value of $\vw$ (a lower $\vw\sim 0.6$, for example, could enhance the SNR significantly~\cite{Braathen:2025svl}) and on the threshold in the SNR above which signals will actually be reconstructed.\footnote{The actual observability of a signal of stochastic GW will depend on various factors --- including the detector noise, foreground modelling, the observation time and the reconstruction method --- several of which are not settled at present. The threshold condition of $\mathrm{SNR}>10$ that we use in this work, following the prescription of the LISA collaboration~\cite{Caprini:2019egz,LISACosmologyWorkingGroup:2022jok}, should be understood as a simplified criterion aimed at circumventing a detailed discussion of the GW signal reconstruction (that would go beyond the scope of this work), rather than a strict boundary between observable and non-observable signals.} However, even in the most optimistic case, a detection of primordial GW from the SFOEWPT would only be possible for a borderline region of BP2. We note that for BP2, the phase transition parameters obtained with \texttt{BSMPTv3} are in the ranges $\alpha\in[0,0.15]$, $\beta/H_\star\in[0,15000]$ and $T_{*}\in[50,135]$ GeV. 

The right panel of \cref{fig:BP2results} shows for the same scan points the $\kappa_\lambda^{(1)}$ value%
\footnote{
We denote, wherever relevant, the tree-level (one-loop corrected) value as $\kappa_\lambda^{(0)}$ ($\kappa_\lambda^{(1)}$).
}%
~as a function of $m_H$, while the colour coding additionally indicate the value of $\xi_n$. Unlike BP1, large BSM deviations in $\kappa_\lambda^{(1)}$ are found for all points exhibiting a SFOEWPT (i.e.\ with $\xi_n>1$), with a minimal value of $\kappa_\lambda^{(1)}\gtrsim 1.35$ (while at tree level one finds $\kappa_\lambda^{(0)}\gtrsim 1.45$). Moreover, the strength of the SFOEWPT increases with 
$\kappa_\lambda$ --- corresponding to the fact that a larger barrier in the potential along the doublet-field direction leads to a stronger first-order EWPT, until a maximal value above which vacuum trapping occurs. These results are similar to those of Ref.~\cite{Biekotter:2022kgf} for the 2HDM, and in particular the range of $1.5\lesssim \kappa_\lambda^{(1)}\lesssim 2.2$ found therein for points with a SFOEWPT in the 2HDM. Finding significant BSM deviations in the trilinear Higgs self-coupling provides motivation to explore the phenomenology of points from BP2 at current and future colliders, and the possibility to reconstruct the shape of the scalar potential realised in this type of scenarios. 

BSM deviations in $\kappa_\lambda$, within the range $1.4 \lesssim \kappa_\lambda\lesssim 2$ like what is found for points in BP2, typically imply a decrease in the cross-section of the di-Higgs production process at the (HL-)LHC, unless resonant contributions or interference therewith compensate the effect from the change in $\kappa_\lambda$ alone. On the other hand, the cross-section for the process $e^+e^-\to Zhh$, which dominates at $e^+e^-$ colliders up to energies somewhat above 1 TeV, grows monotonically with $\kappa_\lambda$, and enhancement with respect to the SM prediction can be expected. 
For this reason as well as for brevity, we do not discuss here the case of Higgs pair production at the HL-LHC, and we refer the interested reader to Ref.~\cite{Braathen:2025svl} for this part of our analysis. We focus instead on Higgs pair production processes at a high-energy $e^+e^-$ collider, considering the case of ILC1000. 

\begin{figure}[ht!]
    \centering
    \includegraphics[width=0.49\textwidth]{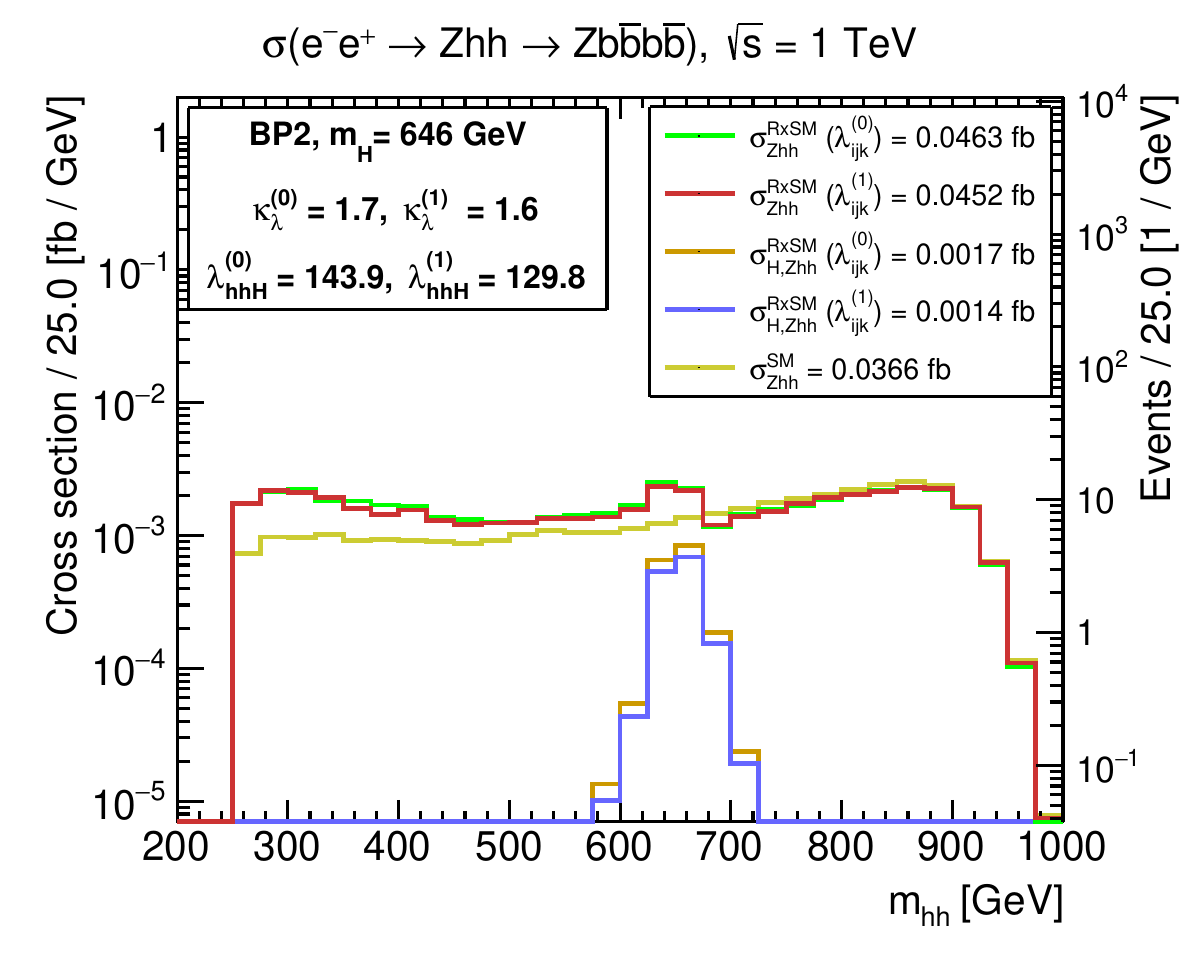}
    \includegraphics[width=0.49\textwidth]{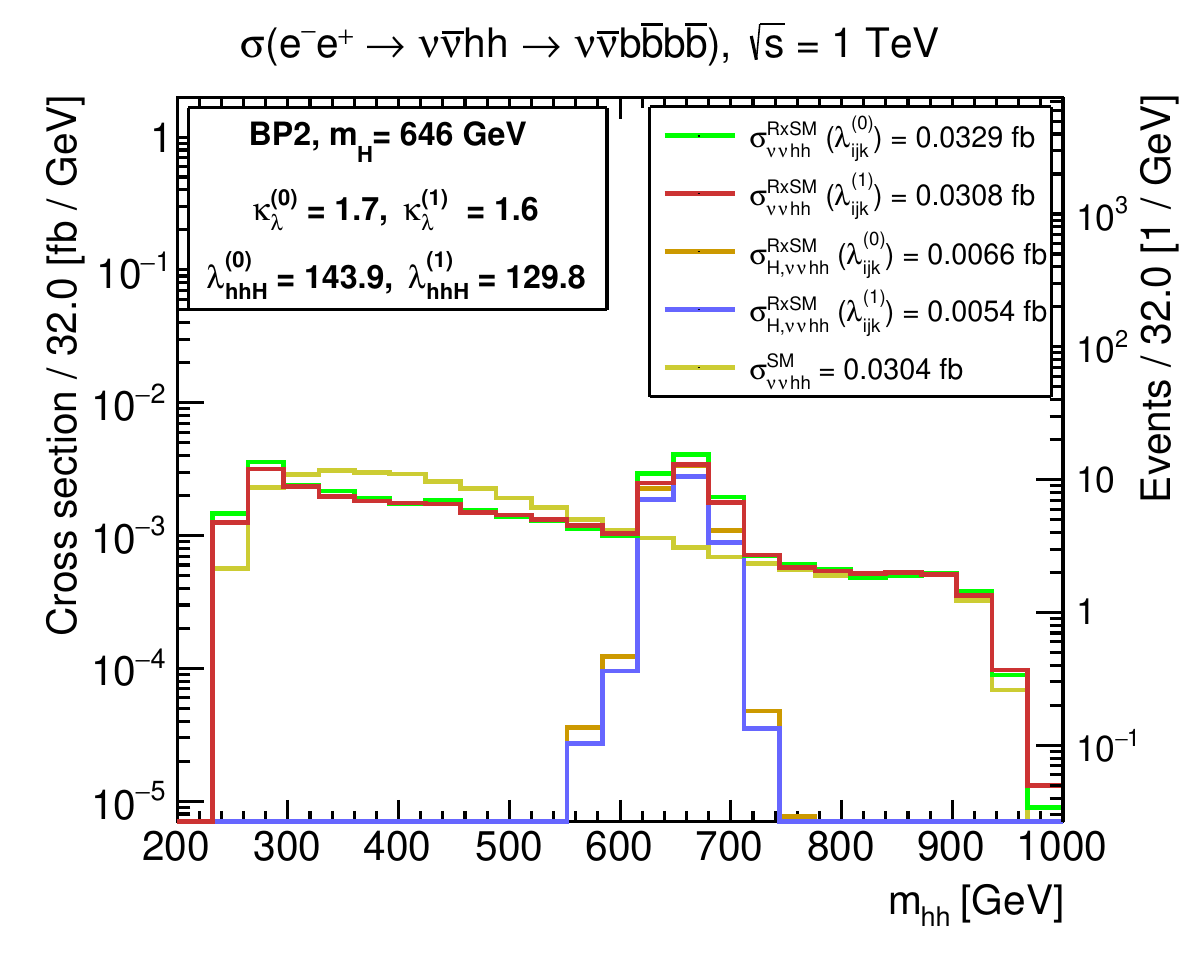}
    \caption{Differential polarised di-Higgs production cross-section distributions as a function of the di-Higgs invariant mass $m_{hh}$, at the ILC1000. \textit{Left:} for the process $e^+e^- \to Zhh \to Zb\bar{b}b\bar{b}$; \textit{right:} for the process $e^+e^-\to \nu\bar\nu hh\to \nu\bar\nu b\bar bb\bar b$. The right axis indicates the sum of event numbers for the two polarisations $(\mp 80\%,\pm 30\%)$  of the electrons and positrons, considering $\cLint = 3200 \text{ fb}^{-1}$ each.
    Values of $\lahhH$ are shown in GeV.}
    \label{fig:mhhdistrib}
\end{figure}

The two panels of \cref{fig:mhhdistrib} display differential distributions with respect to the di-Higgs invariant mass, $m_{hh}$, for the processes $e^+e^-\to Zhh$ (left) and $e^+e^-\to \nu\bar\nu hh$ (right), furthermore taking into account in both cases the $b\bar b$ decay channel of the Higgs bosons. The distributions are obtained for a point from benchmark plane 2 (c.f.\ \cref{eq:defBP2}) characterised by $m_H=646\text{ GeV}$ and $\kappa_{SH}=-1429 \gev$, for which we find a SFOEWPT with $\xi_n=3.9$. Moreover, results are shown for a $(-80\%,+30\%)$ polarisation  of the initial electrons and positrons assuming $\cLint = 3200 \text{ fb}^{-1}$, and after applying cuts and smearing (see \citeres{Arco:2025nii,Braathen:2025qxf,Braathen:2025svl} for further details on the analysis setup). Green and red lines correspond to the full process distributions in the RxSM, using respectively tree-level and one-loop corrected values for the trilinear scalar couplings (we denote these $\lambda_{ijk}^{(0)}$ and $\hat\lambda_{ijk}^{(1)}$, respectively). The yellow lines are the full SM distribution (for the SM, the difference between using $\lambda_{ijk}^{(0)}$ or $\hat\lambda_{ijk}^{(1)}$ would not be distinguishable in these plots). The orange (purple) lines represent the resonant $H$ contribution alone, using $\lambda_{ijk}^{(0)}$ ($\hat\lambda_{ijk}^{(1)}$). Based on the approach of Ref.~\cite{Arco:2025nii}, we defined statistical significances for \textit{(i)} distinguishing the RxSM distributions from the SM and \textit{(ii)} distinguishing the resonant peak from the continuum. 
For both processes, evidence for the resonant contribution can be obtained, even after experimental effects and statistical uncertainties are taken into account, and in the case of $e^+e^-\to \nu\bar\nu hh$ (right) significances of more than $5\sigma$ are achieved. Away from the resonance, the RxSM distributions can also be distinguished from the SM one --- with an overall increase in the case of $e^+e^-\to Zhh$ and a decrease for $e^+e^-\to \nu\bar\nu hh$, which originates from the (constructive or destructive) interference of diagrams containing and not containing $\lambda_{hhh}$. 
Statistical significances of more than $5\sigma$ are obtained for both processes. Lastly, we find that the inclusion of loop corrections to the trilinear scalar coupling leads to a moderate decrease in all computed significances (in other words they make it more difficult to distinguish the RxSM from the SM and to detect the resonant peak), however, not to the point of modifying the qualitative results of the analysis.  

\section{Conclusions}

In this work, we have investigated the RxSM, which is one of the simplest extensions of the SM that can accommodate a SFOEWPT. At the same time, this model already offers a rich phenomenology, depending on whether the behaviour of the EWPT is driven by the doublet or singlet fields of the theory. We have devised two representative benchmark planes for each of these cases, and studied the phenomenology that would be possible at colliders and at LISA. On the one hand, in the benchmark plane dominantly featuring a SFOEWPT driven by the singlet field direction, strong GW signals are found, 
while couplings of the detected Higgs boson are very SM-like, making collider signals challenging. 
On the other hand, for the benchmark plane where the SFOEWPT is driven by the doublet field direction, detectable GW signals are only found for a very narrow part of parameter space. However, the occurrence of a SFOEWPT is in this case correlated with values of the trilinear Higgs self-coupling $1.35 < \kappa_\lambda^{(1)} < 1.7$ (and $1.45 < \kappa_\lambda^{(0)} < 1.9$). This leads to interesting signals at colliders, in particular at high-energy linear $e^+e^-$ colliders where both $e^+e^-\to Zhh$ and $e^+e^-\to \nu\bar\nu hh$ processes would allow discriminating the RxSM from the SM. Moreover, information could be obtained also about the resonant peak --- with the process $e^+e^-\to \nu\bar\nu hh$ being the most promising in this context.  

Our work illustrates the crucial complementarity between different experimental probes, required to access the entire region of the RxSM parameter
space giving rise to SFOEWPTs. 

\subsection*{Acknowledgements}
J.B.\ and A.V.S.\ acknowledge support by the Deutsche Forschungsgemeinschaft (DFG, German Research Foundation) under Germany's Excellence Strategy --- EXC 2121 ``Quantum Universe'' --- 390833306. This work has been partially funded by the Deutsche Forschungsgemeinschaft (DFG, German Research Foundation) --- 491245950. J.B. and A.V.S are supported by the DFG Emmy Noether Grant No.\ BR 6995/1-1. The work of S.H.\ has received financial support from the
grant PID2019-110058GB-C21 funded by
MCIN/AEI/10.13039/501100011033 and by ``ERDF A way of making Europe'', 
and in part by by the grant IFT Centro de Excelencia Severo Ochoa CEX2020-001007-S
funded by MCIN/AEI/10.13039/501100011033. 
S.H.\ also acknowledges support from Grant PID2022-142545NB-C21 funded by
MCIN/AEI/10.13039/501100011033/ FEDER, UE. 
S.H.\ thanks the Particle Theory and Cosmology Group, Center for Theoretical Physics of the Universe at the 
Institute for Basic Science (IBS), Daejeon, South Korea for hospitality during the final stages of this work.

\bibliography{bibliography.bib}

\end{document}